\documentclass[prb,aps]{revtex4}

\begin{document}

\title{Numerical simulation of transmission coefficient using c-number
Langevin equation}
\author{Debashis~Barik$^1$, Bidhan Chandra Bag$^2$
and Deb~Shankar~Ray$^1${\footnote{Email address:
pcdsr@mahendra.iacs.res.in}}}
\affiliation{$^1$Indian Association
for the Cultivation of Science, Jadavpur,
Kolkata 700 032, India\\
$^2$Department of Chemistry, Visva-Bharati, Santiniketan 731 235,
India}

\date{\today}

\begin{abstract}
We numerically implement the reactive flux formalism on the basis
of a recently proposed c-number Langevin equation [Barik
\textit{et al}, J. Chem. Phys. {\bf 119}, 680 (2003); Banerjee
\textit{et al}, Phys. Rev. E {\bf 65}, 021109 (2002)] to calculate
transmission coefficient. The Kramers' turnover, the $T^2$
enhancement of the rate at low temperatures and other related
features of temporal behaviour of the transmission coefficient
over a range of temperature down to absolute zero, noise
correlation and friction are examined for a double well potential
and compared with other known results. This simple method is
based on canonical quantization and Wigner quasiclassical phase
space function and takes care of quantum effects due to the
system order by order.

\end{abstract}

\maketitle
\section{{\bf Introduction}}
The classic work of Kramers \cite{kramer} on the diffusion model
of classical reaction laid the foundation of the modern dynamical
theory of activated processes. Since then the field of chemical
dynamics has grown in various directions to extend the Kramers'
result to non-Markovian friction \cite{grones,car}, to generalize
to higher dimensions \cite{lan,pollak} and to semiclassical and
quantum rate theories
\cite{ray4,weiss,gra5,hangg,woly,mill,calde,topa,bern,db1,skb,db2,db3,barik}
to apply extensively to biological processes \cite{juli} and to
several related issues. An important endeavor in this direction
is the formulation of reactive flux theory
\cite{hangg,barik,keck,san} which has been worked out in detail
over the last two decades by several groups. The method is
essentially based on the realization of rate constant of a
chemical reaction as a Green-Kubo correlation function calculated
at the transition state which acts as a dividing surface between
the reactant and the product states. In the spirit of classical
theory of Kohen and Tannor \cite{tan} we have recently formulated
a quantum phase space function approach to reactive flux theory
to derive a transmission coefficient in the intermediate to
strong damping regime \cite{barik}. The object of the present
paper is threefold: (i) to extend the treatment to numerical
simulation for calculation of time-dependent transmission
coefficient (ii) to analyze the Kramers' turnover, the $T^2$
enhancement of the rate at low temperature and some related
features of temporal behaviour of the time-dependent transmission
coefficient over a wide range of friction, noise correlation and
temperature down to vacuum limit and (iii) to confirm the
validity of our analytical result \cite{barik} in the spatial
diffusion limited regime.

The present scheme of simulation is based on our recently
proposed c-number Langevin equation \cite{db1,skb,db2,db3,barik}
coupled to a set of quantum dispersion equations which take care
of quantum corrections order by order. In what follows we make
use of these equations to follow the evolution of an ensemble of
c-number trajectories starting at the top of the barrier and
sampling only the activated events and recrossing according to
classical numerical reactive flux formulation of  Straub and
Berne \cite{st1,st2,st3}.

The outlay of the paper is as follows: we first give an outline
of c-number Langevin equation in Sec.II followed by a discussion
on the method of numerical simulation for calculation of
transmission coefficient using c-number reactive flux formalism
in Sec.III. The numerical results are compared with those of
others and discussed in Sec.IV. The paper is concluded in Sec.V.

\section{{\bf A brief outline of c-Number Langevin equation
and the coupled quantum dispersion equations.}}

To start with we consider a particle coupled to a medium
consisting of a set of harmonic oscillators with frequency
$\omega_i$. This is described by the following Hamiltonian.

\begin{equation}\label{1}
H = \frac{\hat{p}^2}{2} + V(\hat{q}) +
\sum_j\left\{{\frac{\hat{p}_j^2}{2} +
\frac{1}{2}\kappa_j(\hat{q}_j - \hat{q})^2}\right\}
\end{equation}

Here the masses of the particle and the reservoir oscillators
have been taken to be unity. $\hat{q}$ and $\hat{p}$ are the
coordinate and momentum operators of the particle, respectively
and the set $\{\hat{q}_j,\hat{p}_j\}$ is the set of co-ordinate
and momentum operators for the reservoir oscillators linearly
coupled to the system through coupling constant $\kappa_j$.
$V(\hat{q})$ denotes the external potential field which, in
general, is nonlinear. The coordinate and momentum operators
follow the usual commutation relations $[\hat{q},\hat{p}] =
i\hbar$ and $[\hat{q}_i,\hat{p}_j] = i\hbar\delta_{ij}$.

Eliminating the reservoir degrees of freedom in the usual way we
obtain the operator Langevin equation for the particle

\begin{equation}\label{2}
\ddot{ \hat{q} }(t) + \int_0^t dt' \gamma(t-t') \dot{ \hat{q} }
(t') + V' (\hat{q}) = \hat{F} (t)
\end{equation}

where the noise operator $\hat{F} (t)$ and the memory kernel
$\gamma (t)$ are given by

\begin{equation}\label{3}
\hat{F} (t) = \sum_j \left[ \{ \hat{q}_j (0) - \hat{q} (0)\}
\kappa_j \cos\omega_jt + \hat{p}_j (0) \kappa_j^{1/2}
\sin\omega_jt \right]
\end{equation}
and
\begin{eqnarray*}
\gamma (t) = \sum_j \kappa_j \cos\omega_jt\nonumber
\end{eqnarray*}
or in the continuum limit
\begin{equation}\label{4}
\gamma(t) = \int_0^\infty \kappa(\omega) \rho(\omega) \cos\omega
t \; d\omega
\end{equation}
$\rho (\omega)$ represents the density of the reservoir modes.

Eq.(\ref{2}) is an exact operator Langevin equation for which the
noise properties of $\hat{F}(t)$ can be defined using a suitable
canonical initial distribution of bath co-ordinates and momenta.
In what follows we proceed from this equation to derive a
Langevin equation in c-numbers. The first step towards this goal
is to carry out the {\it quantum mechanical average} of
Eq.(\ref{2}).

\begin{equation}\label{5}
\langle \ddot{ \hat{q} }(t) \rangle + \int_0^t \;dt' \gamma(t-t')
\langle \dot{ \hat{q} }(t') \rangle + \langle V'( \hat{q} )
\rangle = \langle \hat{F} (t) \rangle
\end{equation}

where the average $\langle....\rangle$ is taken over the initial
product separable quantum states of the particle and the bath
oscillators at $t=0$, $|\phi \rangle \{ |\alpha_1\rangle
|\alpha_2\rangle.......|\alpha_N\rangle \}$. Here $|\phi\rangle$
denotes any arbitrary initial state of the particle and
$|\alpha_i\rangle$ corresponds to the initial coherent state of
the i-th bath oscillator. $|\alpha_i\rangle$ is given by
$|\alpha_i \rangle = \exp(-|\alpha_i|^2/2) \sum_{n_i=0}^\infty
(\alpha_i^{n_i} /\sqrt{n_i !} ) | n_i \rangle $, $\alpha_i$ being
expressed in terms of the mean values of the co-ordinate and
momentum of the i-th oscillator, $\langle \hat{q}_i (0) \rangle =
( \sqrt{\hbar} /2\omega_i) (\alpha_i + \alpha_i^\star )$ and
$\langle \hat{p}_i (0) \rangle = i \sqrt{\hbar\omega_i/2 }\;
(\alpha_i^\star - \alpha_i )$, respectively. Now $\langle \hat{F}
(t) \rangle$ is a classical-like noise term which, in general, is
a nonzero number because of the quantum mechanical averaging over
the co-ordinate and momentum operators of the bath oscillators
with respect to initial coherent states and arbitrary initial
state of the particle and is given by

\begin{equation}\label{6}
\langle \hat{F} (t) \rangle = \sum_j \left[ \{ \langle
\hat{q}_j(0)\rangle - \langle \hat{q}(0)\rangle \} \kappa_j \cos
\omega_j t + \langle \hat{p}_j(0) \rangle \kappa_j^{1/2} \sin
\omega_j t \right]
\end{equation}

Now the operator Langevin equation can be written as

\begin{equation}\label{7}
\langle \ddot{ \hat{q} } \rangle + \int_0^t dt'\; \gamma(t-t')
\langle \dot{\hat{q}}(t') \rangle + \langle V'( \hat{q} ) \rangle
= f(t)
\end{equation}

where $\langle \hat{F}(t)\rangle = f(t)$, denotes the quantum
mechanical mean value.

We now turn to the \textit{ensemble averaging}. To realize $f(t)$
as an effective c-number noise we now assume that the momentum
$\langle \hat{p}_j(0) \rangle$ and the co-ordinate $\langle
\hat{q}_j(0)\rangle - \langle \hat{q}(0)\rangle$ of the bath
oscillators are distributed according to a canonical thermal
Wigner distribution for shifted harmonic oscillator \cite{hil} as,

\begin{equation}\label{8}
{\cal P}_j = {\cal N} \exp \left \{ - \frac{  [ \langle \hat{p}_j
(0) \rangle^2 + \kappa_j \left \{ \langle \hat{q}_j (0) \rangle -
\langle \hat{q} (0) \rangle \right \}^2 ] }{ 2 \hbar \omega_j
\left ( \bar{n}_j + \frac{1}{2} \right ) } \right \}
\end{equation}

so that for any quantum mean value ${\cal O}_j ( \langle\hat{p}_j
(0) \rangle, \{ \langle \hat{q}_j (0) \rangle  - \langle \hat{q}
(0) \rangle \} )$, the statistical average $\langle....\rangle_s$
is

\begin{equation}\label{9}
\langle {\cal O}_j \rangle_s = \int {\cal O}_j\; {\cal P}_j\;
d\langle \hat {p}_j (0) \rangle \; d\{ \langle \hat{q}_j(0)
\rangle - \langle \hat{q} (0) \rangle \}
\end{equation}

Here $\bar{n}_j$ indicates the average thermal photon number of
the j-th oscillator at temperature $T$ and is given by
Bose-Einstein distribution $\bar{n}_j=1 / [\exp (\hbar
\omega_j/kT) - 1]$ and ${\cal N} $ is the normalization constant.

The distribution Eq.(\ref{8}) and the definition of the
statistical average over quantum mechanical mean values
Eq.(\ref{9}) imply that $f(t)$ must satisfy

\begin{equation}\label{10}
\langle f (t) \rangle_s = 0
\end{equation}

and

\begin{eqnarray*}
\langle f(t) f(t') \rangle_s = \frac {1} {2} \sum_j \kappa_j \;
\hbar \omega_j  \left ( \coth \frac {\hbar \omega_j } {2 k T}
\right ) \cos \omega_j (t-t')
\end{eqnarray*}

or in the continuum limit

\begin{eqnarray}\label{11}
\langle f(t) f(t') \rangle_s & = & \frac {1} {2} \int_0^\infty
d\omega \;\kappa(\omega)\; \rho(\omega) \;\hbar \omega \left (
\coth \frac {\hbar \omega } {2 k T} \right ) \cos \omega
(t-t')\nonumber\\ & \equiv & c ( t- t' )
\end{eqnarray}

That is, c-number noise $f(t)$ is such that it is zero-centered
and satisfies the standard fluctuation-dissipation relation as
known in the literature. For other details we refer to [16-20].

We now add the force term $V' ( \langle \hat q \rangle )$ on both
sides of the Eq.(\ref{7}) and rearrange it to obtain

\begin{equation}\label{12}
\ddot{q} (t)  + V'(q) + \int_0^t  dt' \; \gamma (t-t')\;
\dot{q}(t')  = f(t) + Q ( t )
\end{equation}

where we put $\langle \hat q (t) \rangle = q (t)$ and $\langle
\dot{ \hat q }(t) \rangle = p (t)$ ; $q(t)$ and $p (t)$ being
quantum mechanical mean values and also

\begin{equation}\label{13}
Q (t) = V' (q) -\langle V' ( \hat {q} ) \rangle
\end{equation}

represents the quantum correction to classical potential.

Eq.(\ref{12}) which is based on the ansatz Eq.(\ref{8}), is
governed by a c-number noise $f(t)$ due to the heat bath,
characterized by Eq.(\ref{10}), Eq.(\ref{11}) and a quantum
correction term $Q(t)$ characteristic of the nonlinearity of the
potential. The canonical thermal Wigner distribution - the ansatz
Eq.(\ref{8}) is always positive definite. It goes over to a pure
state distribution for the ground state of shifted harmonic
oscillator in the vacuum limit, i.e. , at $T=0$ which is again a
well-behaved \cite{hil} distribution. The quantum nature of the
dynamics thus arises from two sources. The first one is due to
the quantum heat bath. The second one is because of nonlinearity
of the system potential as embedded in $Q(t)$ of Eq.(\ref{15}).
To make the description complete for practical calculation one
has to have a recipe for calculation of $Q(t)$
\cite{db1,skb,db2,db3,barik,sm,akp}. For the present purpose we
summarize it as follows:

Referring to the quantum mechanics of the system in the Heisenberg
picture one may write,

\begin{eqnarray}\label{14}
\hat{q} (t) & = & \langle\hat{q} (t) \rangle + \delta\hat{q}
(t)\nonumber\\
\hat{p}(t) & = & \langle\hat{p} (t) \rangle + \delta\hat{p} (t)
\end{eqnarray}

$\delta\hat{q} (t)$ and $\delta\hat{p} (t)$ are the operators
signifying quantum corrections around the corresponding quantum
mechanical mean values $q$ and $p$. By construction $\langle
\delta\hat{q} \rangle = \langle \delta\hat{p} \rangle = 0$ and $[
\delta\hat{q},\delta\hat{p} ] = i\hbar$. Using Eq.(\ref{14}) in
$\langle V' ( \hat {q} ) \rangle$ and a Taylor series expansion
around $\langle\hat{q} \rangle$ it is possible to express $Q (t)$
as

\begin{equation}\label{15}
Q(t) = -\sum_{n \ge 2} \frac{1}{n!} V^{(n+1)} (q)
\langle\delta\hat{q}^n(t)\rangle
\end{equation}

Here $V^{(n)} (q)$ is the n-th derivative of the potential $V (
q)$. To second order $Q (t)$ is given by $Q(t) = -\frac{1}{2}
V^{\prime\prime\prime} (q) \langle \delta \hat{q}^2(t) \rangle$
where $q(t)$ and $\langle \delta \hat{q}^2 (t) \rangle$ can be
obtained as explicit functions of time by solving following set
of approximate coupled equations Eq.(\ref{16}) to Eq.(\ref{18})
together with Eq.(\ref{12})

\begin{eqnarray}
\frac{d}{dt}\langle \delta \hat{q}^2 \rangle & = & \langle \delta
\hat{q} \delta \hat{p} + \delta \hat{p} \delta \hat{q} \rangle\label{16}\\
\frac{d}{dt}\langle \delta \hat{q} \delta \hat{p} + \delta
\hat{p} \delta \hat{q} \rangle & = & 2\langle \delta \hat{p}^2
\rangle -
2V^{\prime\prime}(q) \langle \delta \hat{q}^2 \rangle\label{17}\\
\frac{d}{dt}\langle \delta \hat{p}^2 \rangle & = &
-V^{\prime\prime}(q) \langle \delta \hat{q} \delta \hat{p} +
\delta \hat{p} \delta \hat{q} \rangle\label{18}
\end{eqnarray}

While the above set of equations provide analytic solutions
containing lowest order quantum corrections, the successive
improvement of $Q(t)$ can be achieved by incorporating higher
order contribution due to the potential $V(q)$ and the
dissipation effects on the quantum correction terms. In Appendix A
we have derived the equations for quantum corrections upto fourth
order \cite{sm} as employed in the present numerical scheme. Under
very special circumstances, it has been possible to include
quantum effects to all orders \cite{db3,akp}. The appearance of
$Q(t)$ as a quantum correction term due to nonlinearity of the
system potential and dependence of friction on frequency make the
c-number Langevin equation (Eq.(\ref{12})) distinct from the
earlier equations of Schmid and Eckern \textit{et al}
\cite{scm,eck}. The approach has been recently utilized by us to
derive the quantum analogues \cite{db1,skb,db2,db3} of classical
Kramers, Smoluchowski and diffusion equations with probability
distribution functions of c-number variables. An important
success of the scheme is that these equations of motion for
probability distribution functions do not contain derivatives of
the distribution functions higher than second for nonlinear
potentials ensuring positive definiteness of the distribution
functions. This is in contrast to usual situations \cite{Loui}
where one encounters higher derivatives of Wigner,
Glauber-Sudarshan distribution functions for nonlinear potential
and the positive definiteness is never guaranteed. More recently
the classical Kohen-Tannor formalism of reactive flux has been
extended to quantum domain \cite{barik}. In what follows we
present a numerical simulation of reactive flux using c-number
Langevin dynamics.

\section{{\bf Numerical simulation}}

The numerical solution of Eq.(\ref{12}), along with Eqs.(\ref{A1})
to Eq.(\ref{A3}) is performed according to the following main
steps.

We first briefly outline the method of generation of c-number
noise. Eq.(\ref{11}) is the fluctuation-dissipation relation and
is the key element for generation of c-number noise. $\langle f(t)
f(t') \rangle_s$ is correlation function which is classical in
form but quantum mechanical in content. We now show that c-number
noise $f(t)$ is generated as a superposition of several
Ornstein-Ulhenbeck noise processes. It may be noted that in the
continuum limit $c(t-t')$ is given by

\begin{equation}\label{19}
 c ( t- t' ) = \frac {1} {2} \int_0^\infty d\omega \;\kappa(\omega)\;
\rho(\omega) \;\hbar \omega \left ( \coth \frac {\hbar \omega }
{2 k T} \right ) \cos \omega (t-t')
\end{equation}

In determining the evolution of stochastic dynamics governed by
Eq.(\ref{12}) it is essential to know a priori the form of
$\rho(\omega)\kappa(\omega)$. We assume a Lorentzian distribution
of modes so that

\begin{equation}\label{20}
\kappa (\omega) \rho(\omega) = \frac{2}{\pi}
\left(\frac{\Gamma}{1 + \omega^2 \;\tau_c^2 }\right)
\end{equation}

where $\Gamma$ and $\tau_c$ are the dissipation in the Markovian
limit and correlation time, respectively. Eq.(\ref{20}) when used
in Eq.(\ref{4}) in the continuum limit yields an exponential
memory kernel $\gamma (t) = (\Gamma/\tau_c) e^{-t/\tau_c}$. For a
given set of parameters $\Gamma$ and $\tau_c$ along with
temperature $T$, we first numerically evaluate the integral (19)
as a function of time. In the next step we numerically fit the
correlation function with a superposition of several exponentials,

\begin{equation}\label{21}
c(t-t')=\sum_i
\frac{D_i}{\tau_i}\;\exp\left(\frac{-|t-t'|}{\tau_i}\right),\;\;\;\;\;\;\;\;i=1,2,3...
\end{equation}

The set $D_i$ and $\tau_i$ the constant parameters are thus
known. In Fig.1 we compare the correlation $c(t)$ determined from
the relation (19) with the superposition (21) for three different
temperatures at $kT=1.0$, $0.5$ and $0.1$ for $\Gamma=1.0$ and
$\tau_c=3.0$. The numerical agreement between the two sets of
curves based on Eq.(\ref{19}) and Eq.(\ref{21}) suggests that one
may generate a set of exponentially correlated color noise
variables $\eta_i$ according to

\begin{equation}\label{22}
\dot{\eta}_i=-\frac{\eta_i}{\tau_i}+\frac{1}{\tau_i}\;\xi_i(t)
\end{equation}

where

\begin{equation}\label{23}
\langle \xi_i(t) \rangle =0 \;\;\;\;and\;\;\;\; \langle \xi_i(0)
\xi_j(\tau)\rangle = 2 D_i\; \delta_{ij}\;
\delta(\tau)\;\;\;\;\;\;\; (i=1,2,3....)
\end{equation}

in which $\xi_i(t)$ is a Gaussian white noise obeying
Eq.(\ref{23}), $\tau_i$, $D_i$ being determined from numerical
fit. The noise $\eta_i$ is thus an Ornstein-Ulhenbeck process
with properties.

\begin{equation}\label{24}
\langle \eta_i(t) \rangle =0 \;\;\;\;and\;\;\;\; \langle \eta_i(t)
\eta_j(t')\rangle = \delta_{ij}\;
\frac{D_i}{\tau_i}\;\exp\left(\frac{-|t-t'|}{\tau_i}\right)
\;\;\;\;\;\;\; (i=1,2,3....)
\end{equation}

Clearly $\tau_i$ and $D_i$ are the correlation time and strength
of the color noise variable $\eta_i$. The c-number noise $f(t)$
due to heat bath is therefore given by

\begin{equation}\label{25}
f(t)=\sum_{i=1}^n \;\eta_i
\end{equation}

Having obtained the scheme for generation of c-number noise
$f(t)$ we now proceed to solve the stochastic differential
equations.

In order to solve the c-number Langevin equation Eq.(\ref{12}) we
may write it in the equivalent form \cite{st3,zwa,st4}

\begin{eqnarray}
\dot{q} & = & p\nonumber\\
\dot{p} & = & -V'(q)+Q(t)+\sum_i\eta_i(t)+z\nonumber\\
\dot{z} & = & -\Gamma
\frac{p}{\tau_c}-\frac{z}{\tau_c}\label{26}\\
\dot{\eta}_i & = &
-\frac{\eta_i}{\tau_i}+\frac{1}{\tau_i}\;\xi_i(t)\nonumber
\end{eqnarray}

The integration of the above set of equations \cite{san,st3} is
carried out using the second order Heun's algorithm. A very small
time step size, $0.001$, has been used.

The above set of equations differ from the corresponding
classical equations in two ways. First, the noise correlation of
c-number heat bath variables $f(t)$ are quantum mechanical in
character which is reflected in $D_i$ and $\tau_i$. Second, the
knowledge of $Q(t)$ requires the quantum correction equations
given in the Appendix A which provides quantum dispersion about
the quantum mechanical mean values $q$ and $p$ of the system. It
is thus essential to take care of these contributions. A simple
way to achieve this is to solve the equations Eq.(\ref{A1}) to
Eq.(\ref{A3}) in the Appendix A for $\langle \delta
\hat{q}^n(t)\rangle$ by starting with N-particles (say, around
5000) all of them above the barrier top at $q=0$, half with a
positive velocity distributed according to velocity distribution
\cite{barik} $p\;\exp\left(\frac{-p^2}{2\hbar \omega_0
(\overline{n}_0+1/2)}\right)$ and the other half with the same
distribution with negative velocities, initial values of
dispersion being set as $\langle \delta
\hat{q}^2(t)\rangle_{t=0}=0.5$, $\langle
\delta\hat{q}\;\delta\hat{p}+\delta\hat{p}\;\delta\hat{q}\rangle_{t=0}=1.0$,
$\langle \delta \hat{p}^2(t)\rangle_{t=0}=0.5$ and with others
set as zero. The width of the distribution is the same as that
for Eq.(\ref{8}) where $\omega_j$ is replaced by $\omega_0$
corresponding to the reactant harmonic well which is in
equilibrium with the bath. We take the time averaged contribution
of the quantum corrections upto $1/\Gamma$ time for each
trajectory to solve the Langevin equation Eq.(\ref{26}) for
N-particles. The time dependent transmission coefficient is
calculated from these sets of simulated data by calculating

\begin{equation}\label{27}
\kappa(t)=\frac{N_+(t)}{N_+(0)}-\frac{N_-(t)}{N_-(0)}
\end{equation}

where $N_+(t)$ and $N_-(t)$ are the particles that started with
positive velocities and negative velocities, respectively and at
time $t$ are in or over the right hand well (i.e. the particles
for which the quantum mean value $q(t)>0$).

\section{{\bf Results and discussions}}

We now consider the potential of the form $V(q)=a\;q^4-b\;q^2$,
where $a$ and $b$ are the two parameters. The other three input
parameters for our calculations are temperature $T$, strength of
noise correlation $\Gamma$ and correlation time $\tau_c$. For the
present purpose we fix $a=0.001$ and $b=0.5$ for the entire
calculation except those for Fig.3 and Fig.8. In order to ensure
the stability of the algorithm we have kept $\Delta{t}/\tau_c<<1$
where $\Delta{t}$ is the integration time step size. Fig.2
exhibits the temporal variation of classical transmission
coefficient (dotted line) and transmission coefficient calculated
by present method (solid line) for two typical different parameter
regimes (a) $\Gamma=3.0$ and (b) $\Gamma=5.0$ for $kT=0.5$,
$\tau_c=3.0$ to illustrate the differential behaviour of the
classical \cite{san} and quantum effects. In the both cases one
observes significant increase in transmission coefficient due to
quantum contribution over and above the magnitude of classical
transmission coefficient. In order to extract out the contribution
of quantum correction due to nonlinearity of the system potential
$Q(t)$ we exhibit in Fig.3 the time dependent transmission
coefficient for the parameter set $a=0.005$, $b=0.5$, $kT=0.025$,
$\Gamma=1.0$ and $\tau_c=1.0$ with (dotted line) and without
(solid line) second order quantum corrections. It is observed
that the quantum corrections due to nonlinearity affects the
stationary values more than the transient ones.

In order to check the workability of the method we now compare in
Fig.4 the temporal variation of numerical (dotted line) and
analytical \cite{barik} (solid line) transmission coefficient at
$kT=1.0$ for two different parameter regimes characteristic of
adiabatic regime (a) $\Gamma=2.0$, $\tau_c=5.0$ (b) $\Gamma=4.0$,
$\tau_c=8.0$ and caging regime (c) $\Gamma=90.0$, $\tau_c=10.0$,
the two regimes being differentiated according to $\omega_b$,
$\Gamma$ and $\tau_c$; adiabatic $(\omega_b^2-\Gamma/\tau_c>0)$;
caging ($(\omega_b^2-\Gamma/\tau_c<0)$, $\omega_b$ refers to the
barrier frequency). We observe that while in the adiabatic regime
the agreement is excellent the numerical transmission coefficients
tend to settle down around zero relatively earlier compared to
analytical one in the caging regime; the numerically calculated
phase of oscillation, however, corresponds correctly to its
analytical counterpart. Keeping in view of the fact that the
analytical results are based on a phase space function approach
which is independent of the method of numerical simulation, we
believe that the agreement is quite satisfactory in both
adiabatic and caging regimes which ensures the validity of the
numerical procedure as followed in the present treatment.

In Fig.5 we show the typical simulation results over a range of
dissipation parameters $\Gamma$ from $0.01$ to $8.0$ for
$\tau_c=1.0$ and $kT=1.0$. For a low value of $\Gamma$, e.g.,
$0.01$ simulation result remains at a high value for some period
to drop rather suddenly and to settle finally in an oscillatory
way to a low asymptotic value. The behaviour is almost
qualitatively same as its classical counterpart as shown by
Sancho, Romero and Lindenberg \cite{san}. The long temporal
oscillation of transmission coefficient calculated by the
c-number method is typical for very low dissipation regime. As
dissipation increases to $0.5$ the temporal variation of
$\kappa(t)$ becomes monotonic and $\kappa(t)$ settles down much
earlier at a much higher asymptotic value. For $\Gamma=1.0$ to
$8.0$ the transmission coefficient decreases rather
significantly. The oscillations at short times for $\Gamma=5.0$
and $8.0$ are due to usual transient recrossings.

The variation of asymptotic transmission coefficient with
dissipation parameter $\Gamma$ in Fig.5 illustrates the well known
turnover phenomenon. In Figs.6-8 we analyze this aspect in greater
detail. To this end we show in Fig.6 how the asymptotic
transmission coefficient as a function of dissipation constant
$\Gamma$ calculated (classical) earlier by Lindenberg \textit{et
al} \cite{san} follows closely to that of ours at relatively high
temperature $kT=3.0$ for $\tau_c=3.0$. The agreement between the
two allows us to have a numerical check on the method and to
ensure the validity of the result in the classical limit. The
turnover problem was investigated earlier by Melnikov \textit{et
al} \cite{melni} and Pollak \textit{et al} \cite{rip} to obtain
correct analytical solution bridging the spatial diffusion and
the energy diffusion regimes and that goes over to the limiting
behaviour at high and low dissipation. To explore the quantum
effects we study the turnover behaviour at various temperatures
down to absolute zero as shown in Fig.7 for $\tau_c=3.0$. As the
temperature is lowered the maximum at which the turnover occurs
shifts to the left and the damping regime that corresponds to
classical energy diffusion, i.e., in the low damping regime
becomes exponentially small as one approaches to absolute zero.
It is fully consistent with the earlier observation \cite{hangg}
made on this issue.

To check the validity of the numerical results on quantum
turnover in a more quantitative way we have further compared in
Fig.8 our results (solid lines) with full quantum results (dotted
lines) of Topaler and Makri \cite{topa} (Fig. 9a-b of Ref.12) for
two different scaled temperatures $kT=1.744 (200 K)$ and
$kT=2.617 (300 K)$ for a double well potential with $a=0.0024$
and $b=0.5$ in the Ohmic regime. It is immediately apparent from
Fig.8 that the transmission coefficients calculated by simulation
of the c-number Langevin equation with fourth order quantum
corrections compared satisfactorily with the full quantum results
based on path integral Monte Carlo method of Topaler and Makri
\cite{topa}.

In Fig.9 we plot the temporal behaviour of transmission
coefficient $\kappa(t)$ for different temperatures for
$\Gamma=1.0$ and $\tau_c=1.0$. It is observed that $\kappa(t)$
quickly settles after a fast fall and the asymptotic values of
the transmission coefficient converge to a temperature
independent value as the temperature is increased from $kT=0.0$
to $kT=5.0$. Fig.10 exhibits this variation of asymptotic value
of transmission coefficient (dotted line) explicitly as a function
of temperature and a comparison with analytical \cite{barik}
results (solid line) for $\Gamma=2.0$ and $\tau_c=5.0$. The
agreement is found to be excellent. In order to analyze the
temperature dependence of the transmission coefficient at low
temperature and compare with earlier results \cite{han}, the
numerical result (solid line) is fitted against a function of the
form (dotted line) $A \exp(B/T^2)$, in the inset of Fig.10, where
A and B are fitting constants. The well known $T^2$ enhancement
of the quantum rate is observed upto a low temperature beyond
which the fitting curve tends to diverge rapidly as
$kT\rightarrow 0$. This divergence of the rate at very low
temperature had been noted earlier in the analytical result of
Grabert \textit{et at} \cite{han}. In the present calculation,
however the transmission coefficient reaches its maximum finite
value within unity and its validity is retained even in the
vacuum limit. Thus in contrast to classical transmission
coefficient the temperature dependence remains a hallmark of
quantum signature of the transmission coefficient. In Fig.11 we
show the temporal behaviour of the transmission coefficient for
several values of the noise correlation time $\tau_c$ for
$kT=1.0$ $\Gamma=1.0$. The asymptotic transmission coefficient
increases as expected from theoretical point of view.

\section{{\bf conclusion}}

The primary aim of this paper is to extend our recent treatment of
c-number Langevin equation to calculate numerically the time
dependent transmission coefficient within the framework of
reactive flux formalism. Since the quantum dynamics is amenable
to a theoretical description in terms of ordinary coupled
equations which are classical looking in form it is possible to
employ the numerical simulation scheme of Straub and Berne in a
quite straight forward way for calculation of transmission
coefficient. There are two special advantages in the procedure.
First, since we are concerned with the dynamics of quantum
mechanical mean values coupled to quantum correction equations,
we need not calculate higher moments $\langle \hat{q}^2 \rangle$
or $\langle \hat{p}^2 \rangle$ etc in any stage of calculations.
This makes the calculation simpler. Secondly, the treatment can
be readily used to calculate the dynamics even in the vacuum
limit, i.e., $kT\rightarrow 0$, where it is expected that because
of the oscillating nature of the real time propagator in path
integral method Monte Carlo schemes pose very serious problems
from applicational point of view. We have extended the classical
simulation procedure to a quantum domain taking into
consideration of the quantum effects in two different ways. The
quantum effects enter through the correlation function of the
c-number noise variables of the heat bath and furthermore,
through nonlinearity of the system potential which is entangled
with quantum dispersion around the quantum mechanical mean values
of the system operators. We summarize the main results as follows:

(i) The present method is a direct extension of classical
simulation method of  Straub and Berne to quantum domain for
calculation of transmission coefficient within a c-number version
of reactive flux formalism. Although the quantum effects due to
heat bath can be taken into account in terms of noise correlation
expressed in quantum fluctuation-dissipation relation, the
quantum dispersion around the quantum mean values of the system
operators are to be calculated order by order. Notwithstanding
the latter consideration the method is efficient when compared to
full quantum mechanical calculation as demonstrated in the
present simulation.

(ii) We have calculated the time dependent transmission
coefficient for a double well potential over a wide range of
friction and temperature and shown that our numerical simulation
results on turnover phenomena, low temperature enhancement of
quantum rate and other features agree satisfactorily with those
calculated analytically/otherwise using phase space function and
other approaches. The differential behaviour of the classical and
transmission coefficients calculated by the classical and present
c-number method has been analyzed in detail. The procedure is
equipped to deal with arbitrary noise correlation, strength of
dissipation and temperature down to vacuum limit, i.e.,
$kT\rightarrow 0$.

(iii) The present approach is independent of path integral
approaches and is based on canonical quantization and Wigner
quasiclassical phase space function and takes into account of the
quantum effects upto a significant degree of accuracy. This
procedure simply depends on the solutions of coupled ordinary
differential equations rather than the multi-dimensional path
integral Monte Carlo techniques and is therefore complementary to
these approaches, much simple to handle and corresponds more
closely to classical procedure.

{\bf Acknowledgement} We are thankful to S. K. Banik for
discussions. The authors are indebted to the Council of Scientific
and Industrial Research for partial financial support under Grant
No. 01/(1740)/02/EMR-II.

\appendix

\begin{appendix}
\section{}

\begin{center}
{\bf Evolution Equations For Higher-Order Quantum Corrections For
Anharmonic Potential}
\end{center}

\noindent The equations upto fourth order for quantum corrections
(corresponding to the contribution of anharmonicity of the
potential) with dissipative effects taking into consideration of
Lorentzian density of reservoir modes of the form Eq.(\ref{20}),
in the limit $\tau_c$ very small, are listed below.

\noindent Equations for the second order are:

\begin{eqnarray}
\frac{d}{dt} \langle \delta \hat{q}^2 \rangle &=& \langle \delta
\hat{q} \delta \hat{p} +
\delta \hat{p} \delta \hat{q} \rangle,                 \nonumber  \\
\frac{d}{dt} \langle \delta \hat{p}^2 \rangle &=& -2\Gamma \langle
\delta \hat{p}^2 \rangle -V^{\prime\prime} \langle \delta \hat{q}
\delta \hat{p} + \delta \hat{p} \delta \hat{q} \rangle -
V^{\prime\prime\prime} \langle \delta \hat{q} \delta \hat{p}
\delta \hat{q} \rangle,
\label{A1}  \\
\frac{d}{dt} \langle \delta \hat{q} \delta \hat{p} + \delta
\hat{p} \delta \hat{q} \rangle &=& -\Gamma \langle \delta \hat{q}
\delta \hat{p} + \delta \hat{p} \delta \hat{q} \rangle 2\langle
\delta \hat{p}^2 \rangle - 2V^{\prime\prime} \langle \delta
\hat{q}^2 \rangle - V^{\prime\prime\prime} \langle \delta
\hat{q}^3 \rangle, \nonumber
\end{eqnarray}

\noindent Those for the third order are:

\begin{eqnarray}
\frac{d}{dt} \langle \delta \hat{q}^3 \rangle &=&
3\langle \delta \hat{q} \delta \hat{p} \delta \hat{q} \rangle,  \nonumber \\
\frac{d}{dt} \langle \delta \hat{p}^3 \rangle &=& -3\Gamma \langle
\delta \hat{p}^3 \rangle -3V^{\prime\prime} \langle \delta \hat{p}
\delta \hat{q} \delta \hat{p} \rangle + V^{\prime\prime\prime}
\left( \frac{3}{2} \langle \delta \hat{q}^2 \rangle \langle \delta
\hat{p}^2 \rangle - \frac{3}{2}
\langle \delta \hat{p} \delta \hat{q}^2 \delta \hat{p} \rangle + \hbar^2 \right), \nonumber \\
\frac{d}{dt} \langle \delta \hat{q} \delta \hat{p} \delta \hat{q}
\rangle &=& -\Gamma \langle \delta \hat{q} \delta \hat{p} \delta
\hat{q} \rangle + 2\langle \delta \hat{p} \delta \hat{q} \delta
\hat{p} \rangle - V^{\prime\prime} \langle \delta \hat{q}^3
\rangle - \frac{V^{\prime\prime\prime}}{2} \left( \langle \delta
\hat{q}^4 \rangle -
{\langle \delta \hat{q}^2 \rangle}^2 \right),                 \label{A2}  \\
\frac{d}{dt} \langle \delta \hat{p} \delta \hat{q} \delta \hat{p}
\rangle &=& -2\Gamma \langle \delta \hat{p} \delta \hat{q} \delta
\hat{p} \rangle + \langle \delta \hat{p}^3 \rangle -
2V^{\prime\prime}
\langle \delta \hat{q} \delta \hat{p} \delta \hat{q} \rangle  \nonumber \\
&+& \frac{V^{\prime\prime\prime}}{2} \left( \langle \delta
\hat{q}^2 \rangle \langle \delta \hat{q} \delta \hat{p} + \delta
\hat{p} \delta \hat{q} \rangle - \langle \delta \hat{q}^3 \delta
\hat{p} + \delta \hat{p} \delta \hat{q}^3 \rangle \right),
\nonumber
\end{eqnarray}

\noindent And lastly, the fourth order equations are:

\begin{eqnarray}
\frac{d}{dt} \langle \delta \hat{q}^4 \rangle &=&
2\langle \delta \hat{q}^3 \delta \hat{p} + \delta \hat{p} \delta \hat{q}^3 \rangle,  \nonumber \\
\frac{d}{dt} \langle \delta \hat{p}^4 \rangle &=& -4\Gamma \langle
\delta \hat{p}^4 \rangle -2V^{\prime\prime} \langle \delta \hat{q}
\delta \hat{p}^3 + \delta \hat{p}^3 \delta \hat{q} \rangle +
2V^{\prime\prime\prime} \langle \delta \hat{q}^2 \rangle \langle
\delta \hat{p}^3 \rangle,
\nonumber  \\
\frac{d}{dt} \langle \delta \hat{q}^3 \delta \hat{p} + \delta
\hat{p} \delta \hat{q}^3 \rangle &=& -\Gamma \langle \delta
\hat{q}^3 \delta \hat{p} + \delta \hat{p} \delta \hat{q}^3 \rangle
-2V^{\prime\prime} \langle \delta \hat{q}^4 \rangle - 3\hbar^2 +
6\langle \delta \hat{p} \delta \hat{q}^2 \delta \hat{p} \rangle   \nonumber   \\
&+& V^{\prime\prime\prime} \langle \delta \hat{q}^2 \rangle
\langle \delta \hat{q}^3 \rangle,
\label{A3}   \\
\frac{d}{dt} \langle \delta \hat{q} \delta \hat{p}^3 + \delta
\hat{p}^3 \delta \hat{q} \rangle &=& -3\Gamma \langle \delta
\hat{q} \delta \hat{p}^3 + \delta \hat{p}^3 \delta \hat{q} \rangle
+ 2\langle \delta \hat{p}^4 \rangle + 3V^{\prime\prime} (\hbar^2 -
2\langle \delta \hat{p} \delta \hat{q}^2 \delta \hat{p} \rangle)   \nonumber  \\
&+& 3V^{\prime\prime\prime} \langle \delta \hat{q}^2 \rangle
\langle \delta \hat{p} \delta \hat{q} \delta \hat{p} \rangle,
\nonumber  \\
\frac{d}{dt} \langle \delta \hat{p} \delta \hat{q}^2 \delta
\hat{p} \rangle &=& -2\Gamma \langle \delta \hat{p} \delta
\hat{q}^2 \delta \hat{p} \rangle -V^{\prime\prime} \langle \delta
\hat{q}^3 \delta \hat{p} + \delta \hat{p} \delta \hat{q}^3 \rangle
+ \langle \delta \hat{p}^3 \delta \hat{q} + \delta \hat{q} \delta \hat{p}^3 \rangle    \nonumber \\
&+& V^{\prime\prime\prime} \langle \delta \hat{q}^2 \rangle
\langle \delta \hat{q} \delta \hat{p} \delta \hat{q} \rangle.
\nonumber
\end{eqnarray}

\noindent The derivatives of $V(q)$, i.e., $V^{\prime\prime}$ or
$V^{\prime\prime\prime}$ etc. in the above expressions are
functions of $q$ the dynamics of which is given by Eq.(\ref{12}).
\end{appendix}

\newpage

\begin{center}
\section*{Figure Captions}
\end{center}

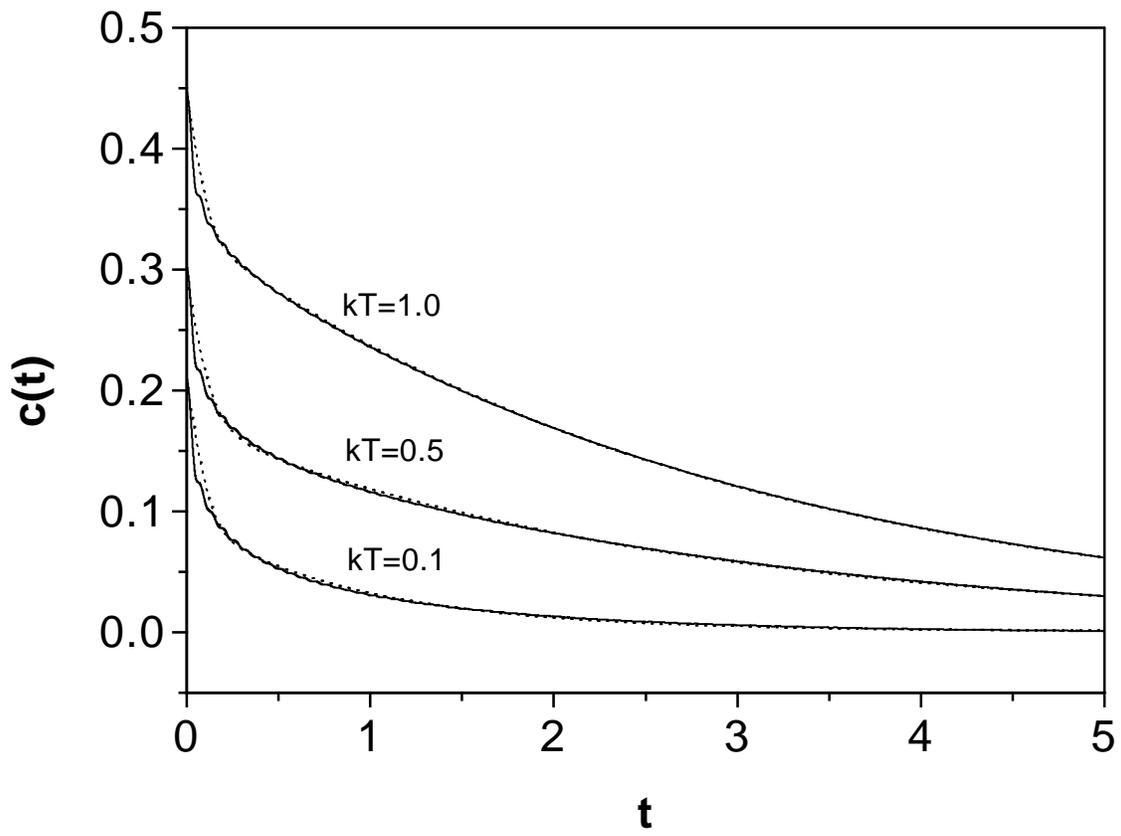
Fig.1: Plot of correlation function $c(t)$ vs $t$ as given by
Eq.(19) (solid line) and Eq.(21) (dotted line) for the set of
parameter values mentioned in the text.

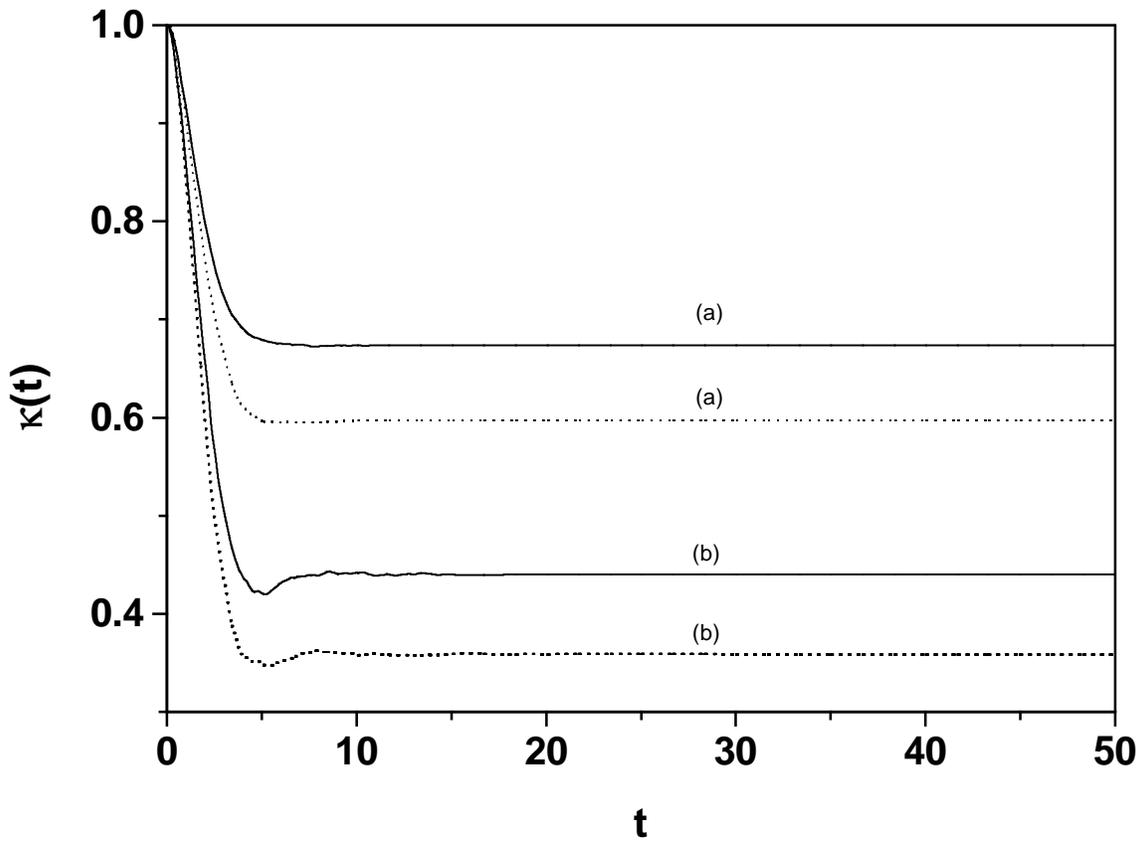
Fig.2: A comparison of transmission coefficients (
classical(dotted line); c-number method(solid line)) as function
of time t is plotted for (a) $\Gamma=3.0$ (b) $\Gamma=5.0$ for the
parameter set mentioned in the text.

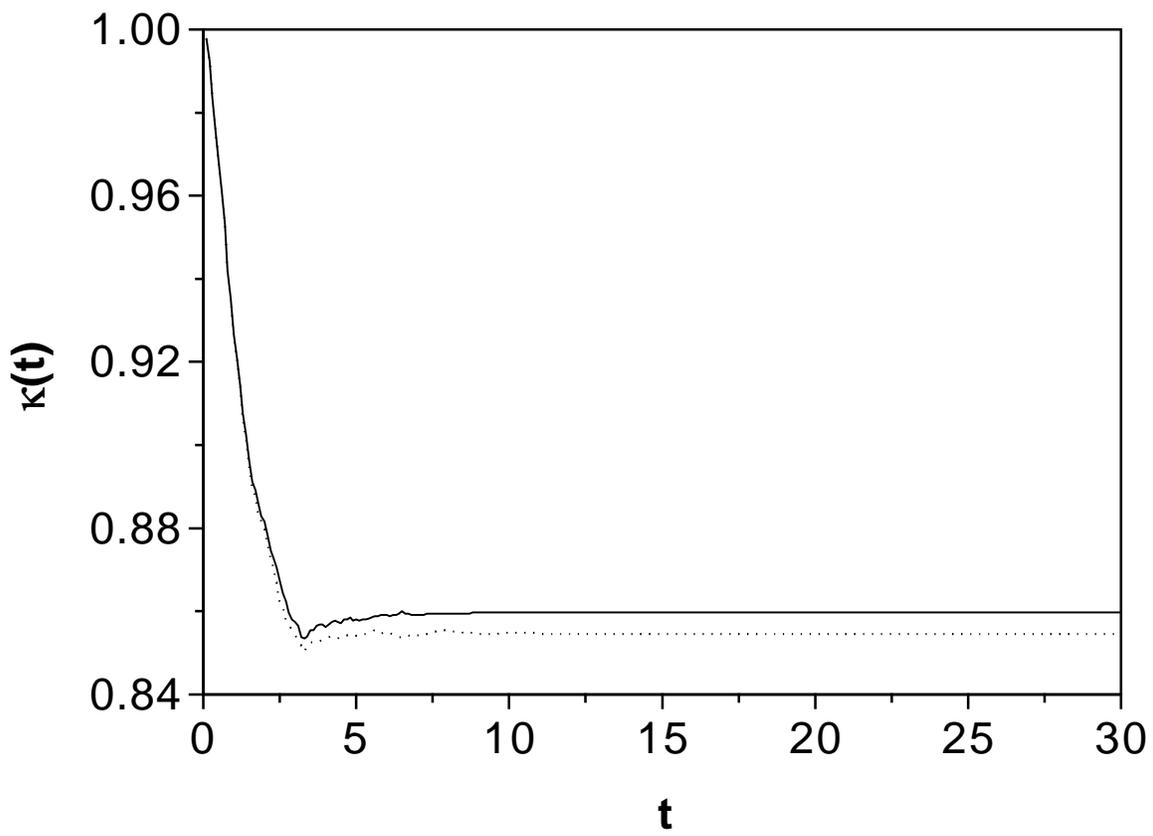
Fig.3: The numerical transmission coefficient $\kappa(t)$ is
plotted against time with (dotted line) and without (solid line)
quantum correction $Q(t)$ for $a=0.005$, $b=0.5$, $\Gamma=1.0$,
$\tau_c=1.0$ at $kT=0.025$

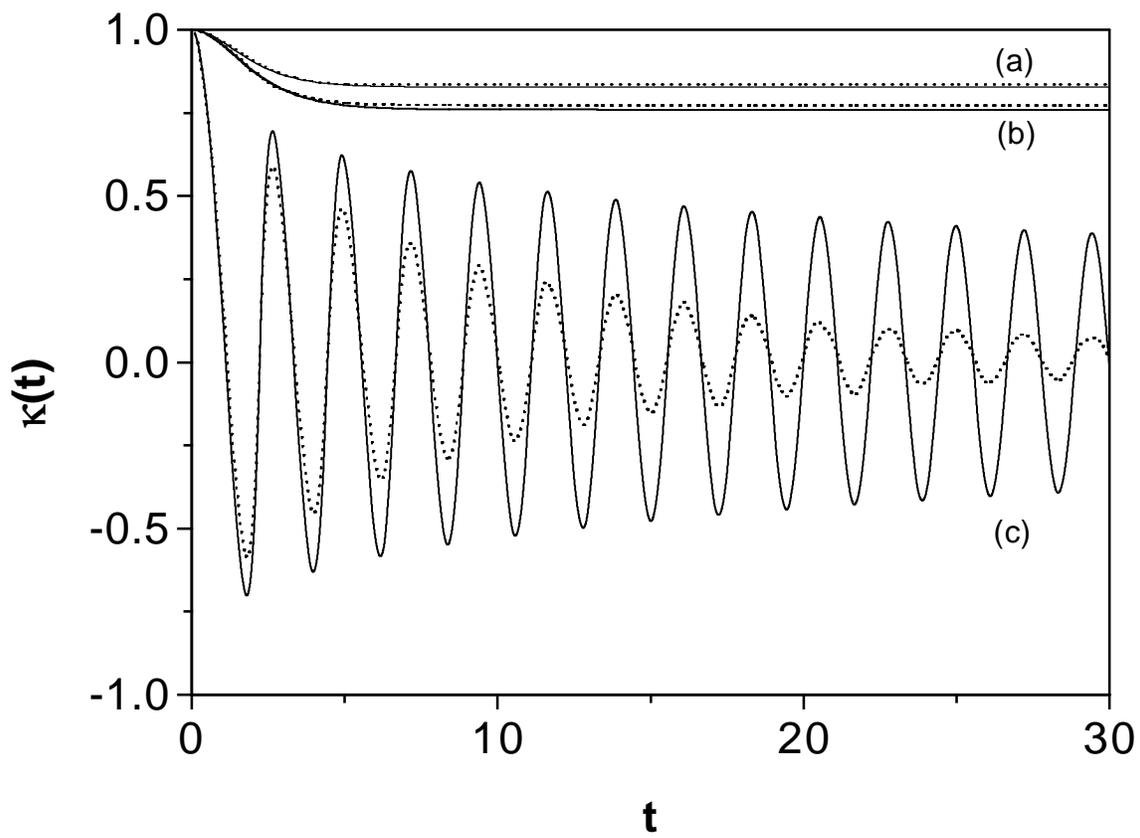
Fig.4: Numerical transmission coefficient $\kappa(t)$ is plotted
against time t and compared with c-number analytical \cite{barik}
$\kappa(t)$ for adiabatic regime [(a) $\Gamma=2.0, \tau_c=5.0$,
(b) $\Gamma=4.0, \tau_c=8.0$] and caging regime [ $\Gamma=90.0,
\tau_c=10.0$] for the parameter set mentioned in the text.

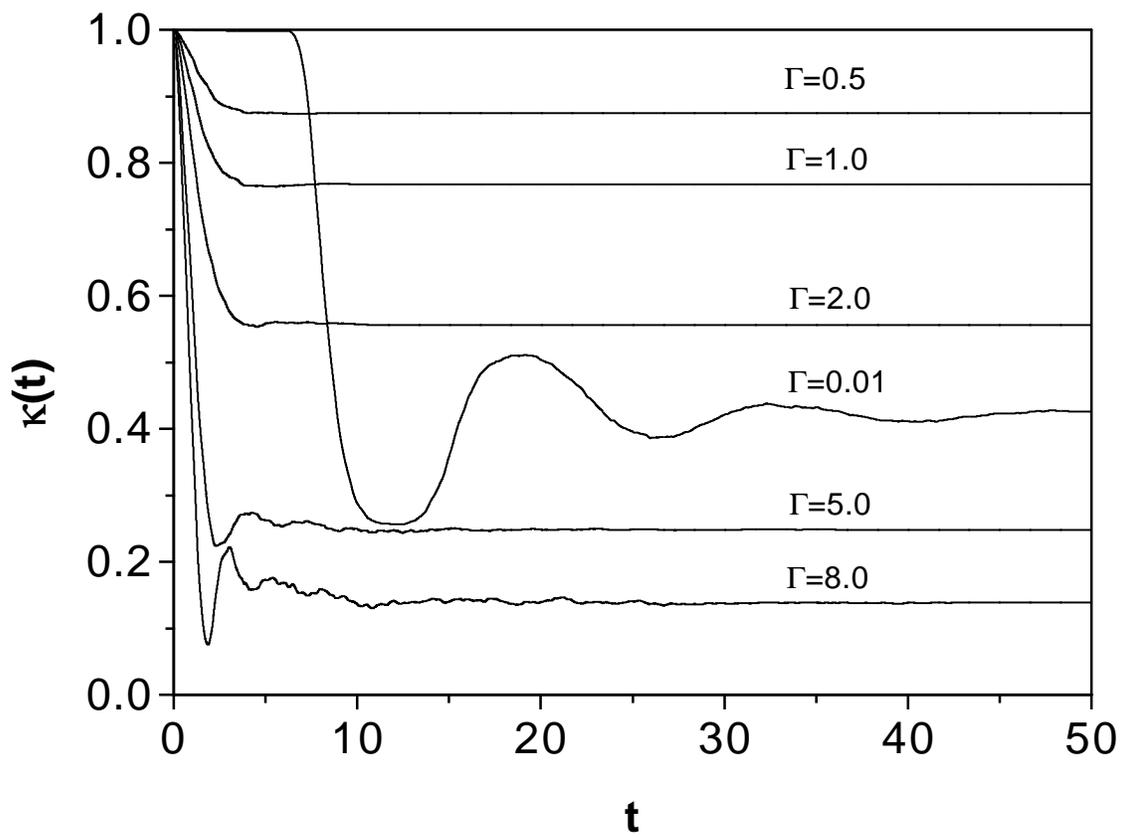
Fig.5: Transmission coefficient, $\kappa(t)$ is plotted against
time for different values of $\Gamma$ for the parameter set
mentioned in the text.

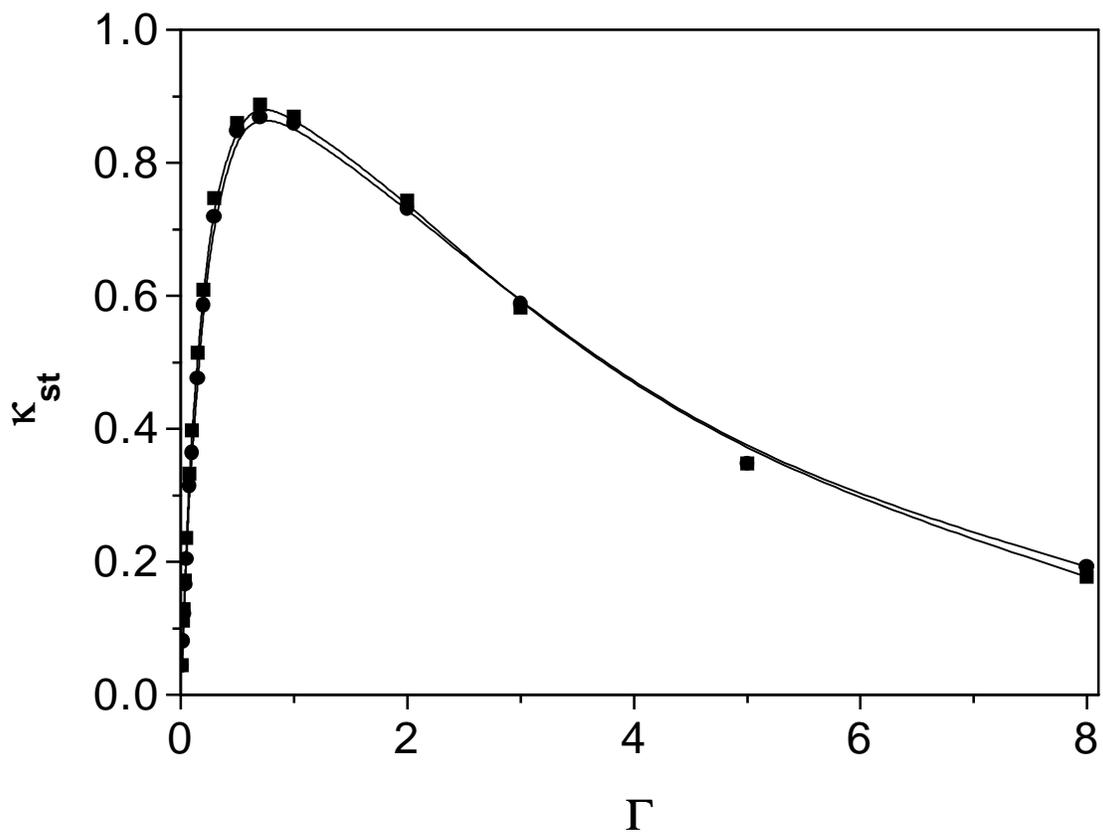
Fig.6: A comparison of the turnover (plot of asymptotic $\kappa$
vs $\Gamma$) calculated by Sancho, Romero and Lindenberg
\cite{san} (circle) with that by the present method (square) for
the parameter set mentioned in the text.

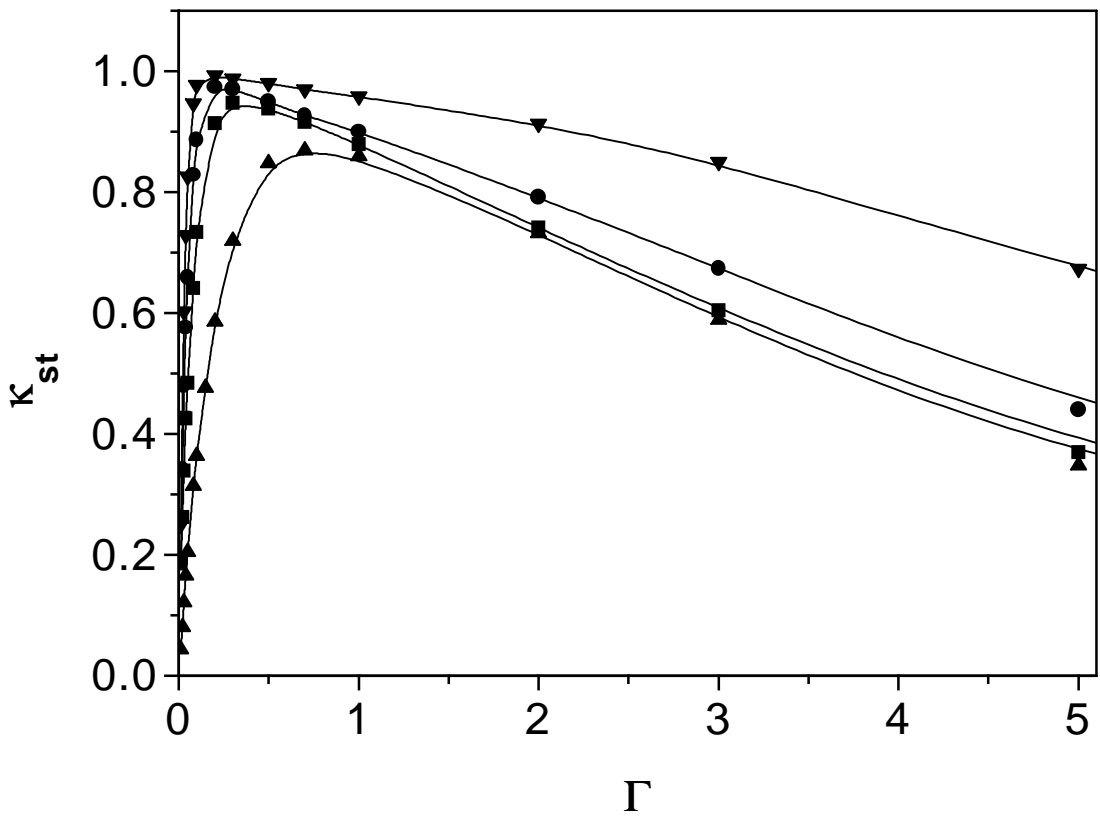
Fig.7: Kramers' turnover (plot of asymptotic $\kappa$ vs $\Gamma$)
for different temperatures for the parameters set mentioned in the
text for $kT=0.0$ (downtriangle), $kT=0.5$ (circle), $kT=1.0$
(square) and $kT=3.0$ (uptriangle).

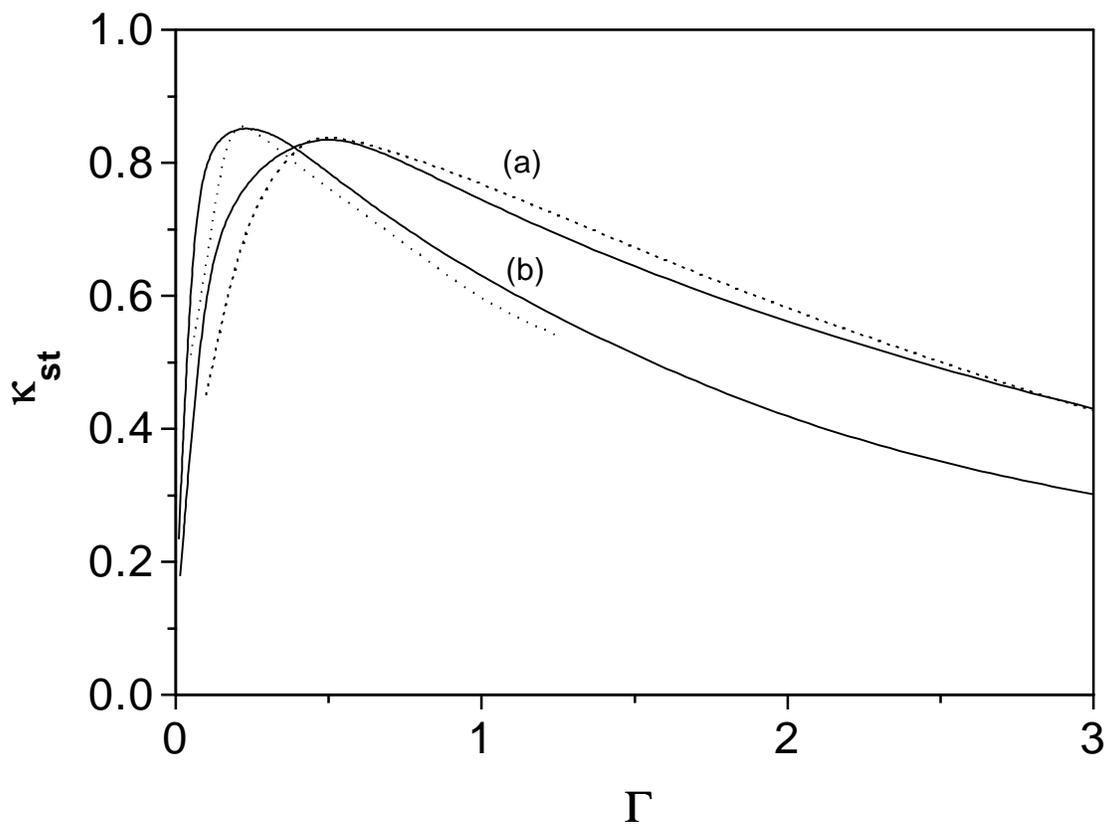
Fig.8: A comparison of the Kramers' turnover (plot of assymptotic
$\kappa$ vs $\Gamma$) for two temperatures (a) $kT=2.617 (300 K)$
and (b) $kT=1.744 (200 K)$ between the present result (solid line)
and full quantum result (dotted line) of Topaler and Makri
(Fig.9a-b of ref.12) for the double well potential with
$a=0.0024$ and $b=0.5$ in the Ohmic regime.

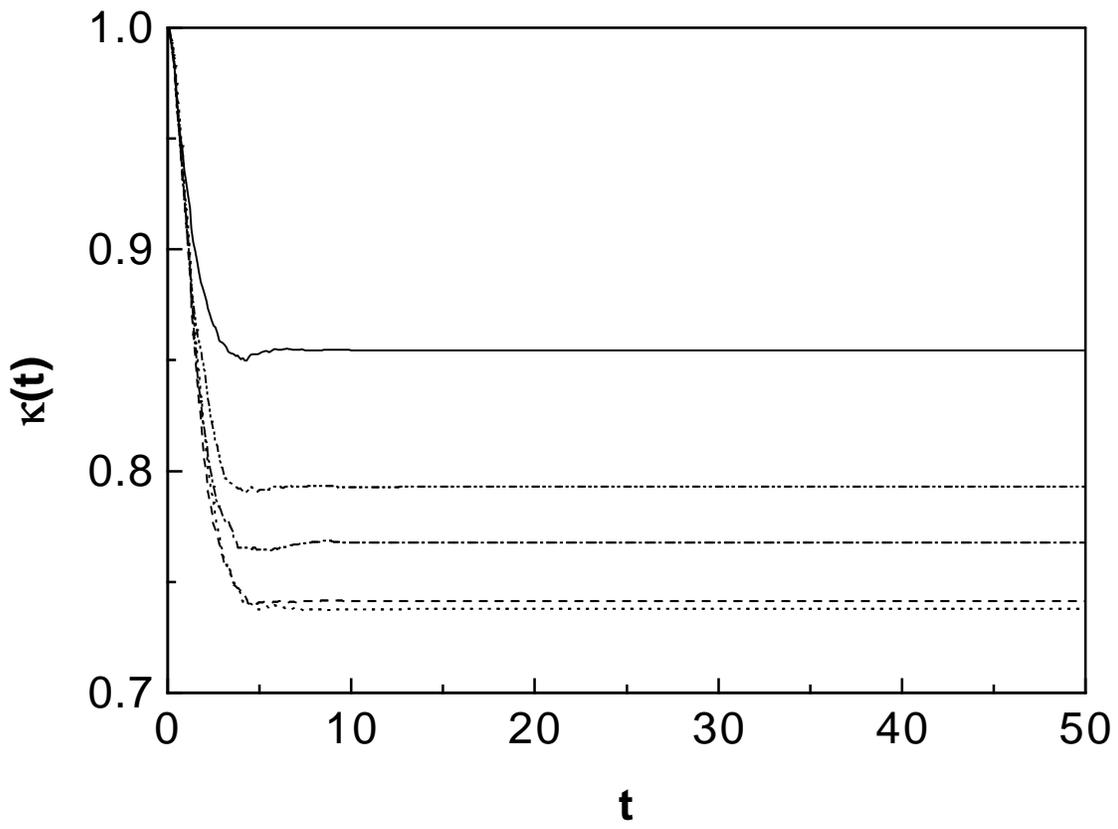
Fig.9: Transmission coefficient $\kappa(t)$ is plotted against
time t for different values of temperature, $kT=0.0$ (solid
line), $kT=0.5$ (dash dot dot line), $kT=1.0$ (dash dot line),
$kT=2.0$ (dashed line) and $kT=5.0$ (dotted line) for the set of
parameters mentioned in the text.

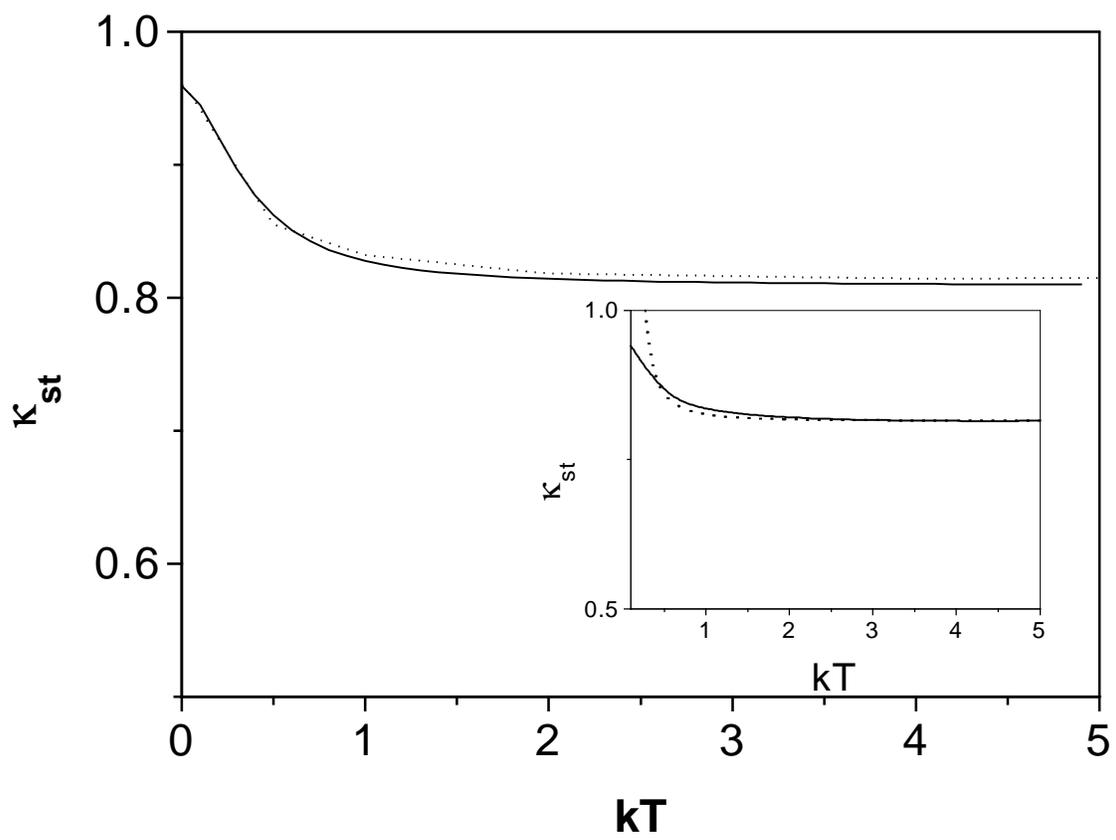
Fig.10: Asymptotic transmission coefficient is plotted against
temperature (dotted line: analytical \cite{barik}, solid line:
numerical) for the parameter set mentioned in the text. Inset:
the same numerical curve (solid line) is plotted against a fitted
curve (dotted) to exhibit $T^2$ enhancement of rate at low
temperature.

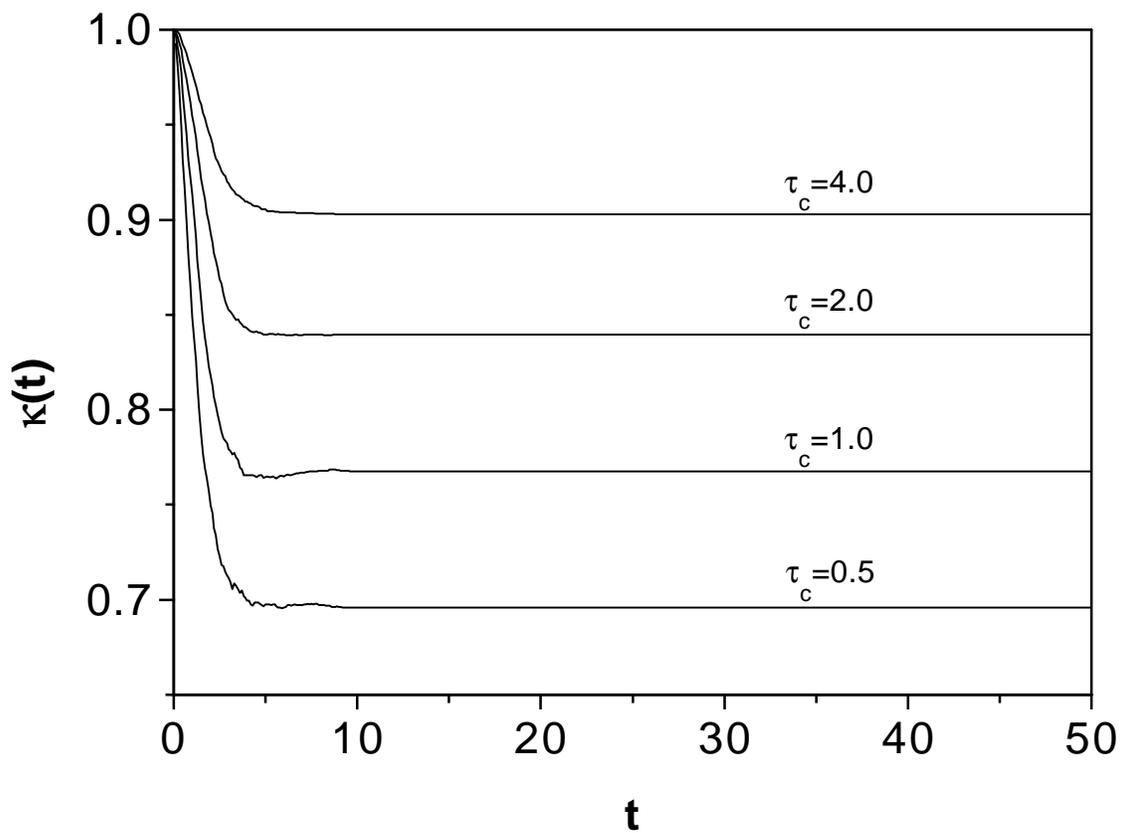
Fig.11: Transmission coefficient $\kappa(t)$ is plotted against
time for different values of $\tau_c$ for the parameter set
mentioned in the text.

\end{document}